\documentclass[twocolumn,showpacs,preprintnumbers]{revtex4} 
\input{epsf}
\usepackage{graphicx}

\def\mev{\mbox{MeV}}

\def\AJ{{\it Astroph. J.} }
\def\AJL{{\it Ap. J. Lett.} }

\def\CQG{{\it Class. Quantum Gravity} }

\def\GRG{{\it Gen. Relativity and Gravitation} }

\def\JHEP{{\it JHEP} }

\def\MPL{{\it Mod. Phys. Lett.} }

\def\NP{{\it Nucl. Phys.} }
\def\PL{{\it Phys. Lett.} }
\def\PR{{\it Phys. Rev.} }
\def\PRL{{\it Phys. Rev. Lett.} }

\def\frac#1#2{{\textstyle{{#1}\over {#2}}}}

\def\lsim{\mathrel{\rlap{\lower4pt\hbox{\hskip1pt$\sim$}}
    \raise1pt\hbox{$<$}}}
\def\gsim{\mathrel{\rlap{\lower4pt\hbox{\hskip1pt$\sim$}}
    \raise1pt\hbox{$>$}}}
\def\sqr#1#2{{\vcenter{\vbox{\hrule height.#2pt
         \hbox{\vrule width.#2pt height#1pt \kern#1pt
         \vrule width.#2pt}
         \hrule height.#2pt}}}}

\def\beq{\begin{equation}}
\def\eeq{\end{equation}}
\def\beqa{\begin{eqnarray}} 
\def\eeqa{\end{eqnarray}}

\begin{document}

\title{WMAP Constraints on  Quintessence}

\author{T. Barreiro}

\altaffiliation[Also at ] {CENTRA, Instituto Superior T\'ecnico, Lisboa.
 Email address: tiago@glencoe.ist.utl.pt}

\author{M. C. Bento}

\altaffiliation[Also at ] {CFIF, Instituto Superior T\'ecnico, Lisboa. 
  Email address: bento@sirius.ist.utl.pt}  

\author{N. M .C. Santos} 

\altaffiliation[Also at ] {CFIF, Instituto Superior T\'ecnico, Lisboa.
 Email address: ncsantos@cfif.ist.utl.pt }

\author{A. A. Sen}

\altaffiliation[Also at ] {CENTRA, Instituto Superior T\'ecnico, Lisboa.
 Email address: anjan@x9.ist.utl.pt} 

\affiliation{ Departamento de F\'\i sica, Instituto Superior T\'ecnico \\
Av. Rovisco Pais 1, 1049-001 Lisboa, Portugal}

\vskip 0.5cm

\date{\today}

\begin{abstract}
     
 We use   recent results from the Wilkinson Microwave Anisotropy Probe (WMAP)
for
the locations of  peaks
 and  troughs of the Cosmic Microwave Background (CMB) power spectrum,
together with  constraints from large-scale structure,  
 to study  a quintessence model
 in which the pure exponential potential is modified by
 a polynomial factor. We find that the model is compatible with all the recent
 data for a wide range of cosmological and potential parameters. Moreover, 
 quintessence is favoured
 compared to
 $\Lambda$CDM for $n_s\approx 1$ and relatively high
 values of the average fraction of dark energy
 before
 last scattering (``early quintessence''); for  $n_s<1$, quintessence and 
$\Lambda$CDM  give similar results, except for high values of 
early quintessence, in which case $\Lambda$CDM is favoured.
\vskip 0.5cm
 
\end{abstract}


\maketitle
\vskip 2pc
\section{Introduction}

Recent cosmological observations suggest that there is a dark energy component
 to the energy density of the universe, which should be added to the matter
 component so as to reach the critical density. Theorists have considered
 various possibilities for the nature of this dark energy, notably 
   a cosmological
 constant and quintessence, a dynamical scalar field
 leading to
 a time-varying equation of state parameter, $w_\phi\equiv p_\phi/\rho_\phi$.
These models most often involve a single field 
\cite{Ratra,Wetterich,Caldwell,Ferreira,Zlatev,Binetruy,Kim,Barreiro,Bento,Uzan}
 or, in some cases, two coupled scalar fields \cite{Fujii,Masiero,Bento1}.
Other possibilities  for the origin of dark energy 
include the generalized Chaplygin gas
 proposal \cite{Chap} and cardassian models \cite{card}.

 In order to unravel the nature of dark energy, it is  crucial
 to use observations so as to be able to discriminate among
 different models. In particular, 
the existence of a dark energy component affects the structure of the CMB power
spectrum, which is particularly sensitive to the amount of dark energy at 
different epochs in cosmology.
 For instance, the
 locations of peaks and troughs 
depend crucially on the amount of dark energy today and at last scattering 
 as well as the dark energy time-averaged equation of state, which are
 model-dependent quantities \cite{Doran1}.  Hence, one can
 use the high precision measurements recently
 obtained by the BOOMERanG \cite{BOOM}, MAXIMA-1 \cite{MAX}, Archeops 
\cite{ARCH} and, in particular,  WMAP \cite{MAP1} observations to
 constrain dark energy models. 

This study has
 already been performed for the case of a cosmological constant \cite{Hu},
 the generalized Chaplygin gas \cite{Chap1}
and for 
some of the quintessence models  that have been
 used in the literature, e.g. for the pure exponential potential
 \cite{Domenico,Doran1},
 a ``leaping kinetic term'' model \cite{Doran1, Doran3}, Ratra-Peebles
 potential 
\cite{Doran1,Doran3,Brax},
 a class of  SUGRA potentials \cite{Brax} and  cosine-type
 quintessence
\cite{Tomo}.

The goal
 of this paper is to study the effect of a dark energy component defined
 by the quintessence potential proposed in 
Ref.~\cite{AS1}

\beq
V(\phi)=\left[A+\left(\phi-\phi_0\right)^2\right] e^{-\lambda \phi}
\label{eq:pot}~,
\eeq
on the location of
 the first three peaks and the first trough of the CMB power spectrum  
 (for a first study  of CMB anisotropies for this model see 
Ref.~\cite{AS2}). We also
 analyze the consequences of
 cluster abundance constraints on $\sigma_8$, the {\it rms} mass
 fluctuation on scales of 8 $h^{-1}$  Mpc.

 The  M-theory motivated potential of Eq.~(\ref{eq:pot}) leads to  a 
 model with some
 interesting features,
 namely there are 
two types of  attractor solutions giving rise  to an accelerating universe
 today,
 corresponding to permanent or transient
 acceleration \cite{Barrow}. Transient vacuum acceleration is a
 particularly appealing scenario that
 would also solve the apparent incompatibility between an  eternally
accelerating universe 
and  string theory, at least in
its present formulation,  given that  string asymptotic states are
 inconsistent with
spacetimes that exhibit event horizons \cite{Hellerman}.

 For both types of solutions,  there is scaling of the densities
 early in the expansion history, with
 $w_\phi=1/3,\ \Omega_\phi\approx 4/\lambda^2$ in the radiation dominated era
 and  $w_\phi=0,\ \Omega_\phi\approx 3/\lambda^2$ in the matter
 dominated epoch, followed by vacuum domination and accelerated expansion.
Transient vacuum dominated solutions  arise for
 $A\lambda^2>1$, in which case the potential has no local minimum or,
 for $A\lambda^2<1$, if the field $\phi$ arrives at the local minimum with
 enough kinetic energy to roll over the potential barrier and continue 
rolling down  the potential. Permanent vacuum domination occurs for
  $A\lambda^2<1$, if $\phi$ gets trapped in the local potential minimum.

Our study of the dependence of  the first three peaks and first
 trough locations
on the potential parameters in the  $(\Omega_m, h, n_s)$ cosmological
 parameters space, in view of WMAP's results, reveals that, 
 for $n_s\approx 1$, the quintessence model of
 Eq.~(\ref{eq:pot}) is  favoured as compared to the $\Lambda$CDM model, 
 provided $\lambda$, which determines 
the average fraction of dark energy before
 last scattering, satisfies $\lambda\lsim 15$.
For $n_s < 1$  and $\lambda\lsim 15$, we find the opposite {\it i.e.} 
 $\Lambda$CDM is favoured as 
compared to quintessence. 
 As $\lambda$ increases, $\lambda\gsim 18$, the model's results  become
 comparable to
 $\Lambda$CDM's, independently of
 $n_s$, as should be expected since the average fraction of dark energy before
 last scattering becomes negligible. 
Moreover, the model presents a
  non-negligible fraction of dark energy at last scattering and during
 structure formation, typical of early quintessence
 models,  leading to suppressed clustering
 power on small length scales as suggested by WMAP/CMB/large scale structure
 combined data \cite{Caldwell2003}. 

\section{Locations of peaks and troughs}

We consider a spatially-flat  Friedmann-Robertson-Walker  (FRW)
Universe containing a perfect  fluid 
with barotropic equation of state $p_w=w\rho_w$, where
 $w$
is a constant ($w=1/3$ for radiation and $w=0$ for dust), together
 with a scalar (quintessence) field  with  potential   given  by
Eq.~(\ref{eq:pot}).   The evolution  equations for  a  spatially-flat FRW
model with Hubble parameter $H\equiv \dot a /a$ are
\beqa
\dot{H} & = &- {1\over2} \left({4\over 3}\rho_r + \rho_m+ \dot\phi^2 \right)~ ,
\nonumber  \\
\dot\rho_r  &=& -4 H\rho_r~,\nonumber     \\
\dot\rho_m  &=& -3 H\rho_m~,\nonumber     \\
\ddot\phi    &=& -3 H\dot\phi-{\partial  V\over \partial \phi}\  ,
\label{eq:Fried}
\eeqa
subject to the Friedmann constraint

\beq
H^2  =  {1\over3} \left(\rho_r+\rho_m +  {1\over2}\dot\phi^2 +V \right) \ ,
\label{Friedcon}
\eeq
We work in units where $M_P\equiv(8 \pi G)^{-1/2}=\hbar =c=1$.

The CMB peaks arise from acoustic oscillations of the primeval plasma just
 before the 
Universe becomes transparent. The angular momentum scale of the oscillations
 is set by the 
acoustic scale $\ell_A$ which for a flat Universe is given by
 \cite{Hu1995,Hu1997}

\beq
\label{eq:la}
\ell_A = \pi {\tau_0 - \tau_{\rm ls} \over \bar c_s \tau_{\rm ls}}~~,
\eeq
where and $\tau=\int  a^{-1}\ dt$ is the conformal time, $\tau_0$ and 
 $\tau_{\rm ls}$
 being, respectively,  the conformal time today
 and at
last scattering;
$\bar{c}_s$ is the average sound speed before decoupling

\beqa
{\bar c}_s &\equiv& \tau_{ls}^{-1} \int_0^{\tau_{ls}} c_s\ d\tau~,\nonumber\\
c_s^{-2}&=&3+{9\over 4}{\rho_b(t)\over\rho_\gamma(t)}~,
\label{eq:bcs}
\eeqa
with $\rho_b/\rho_\gamma$ the ratio of baryon to photon energy density.

\begin{figure*}[ht!]
\begin{center}
 \includegraphics[height=5cm]{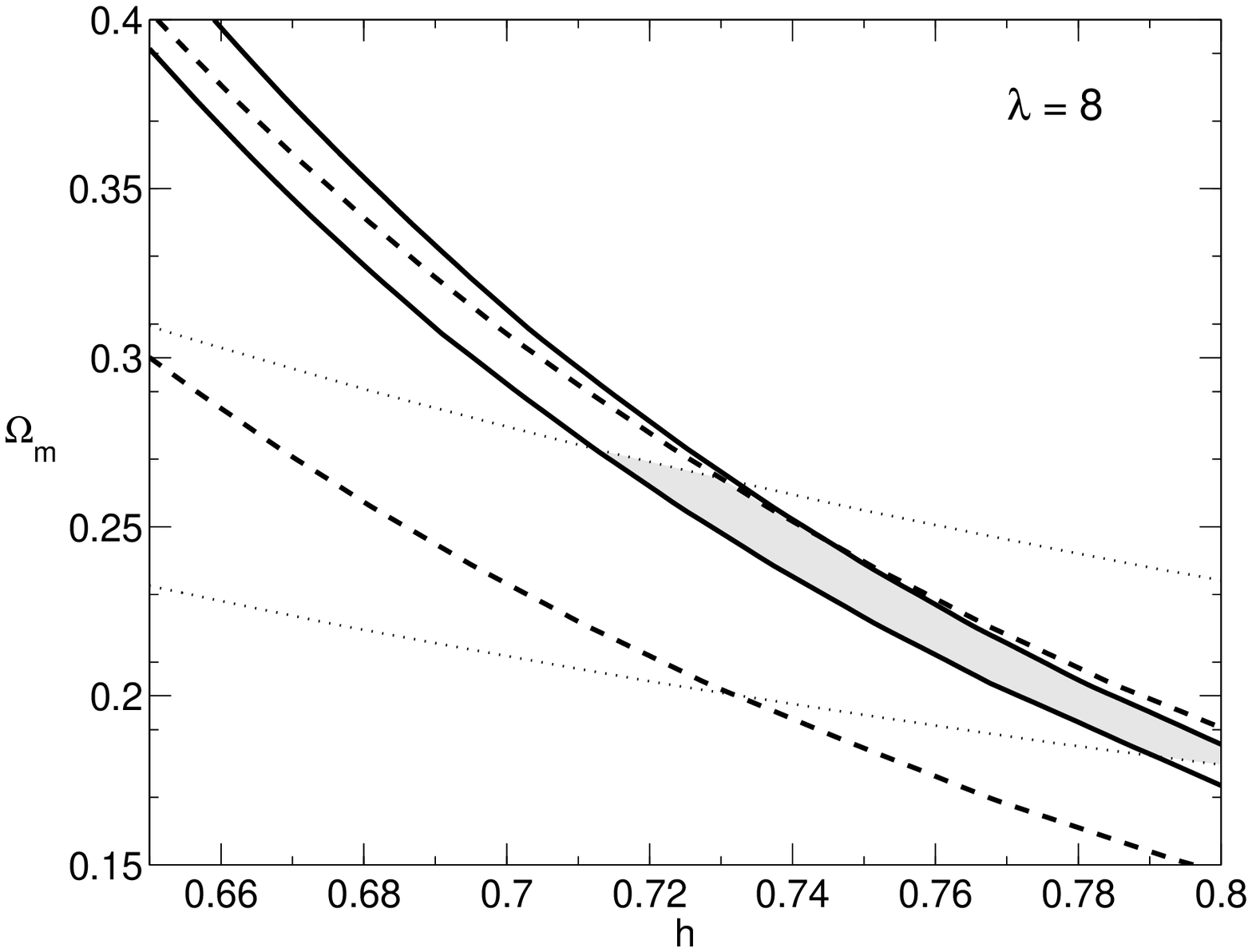}
 \includegraphics[height=5cm]{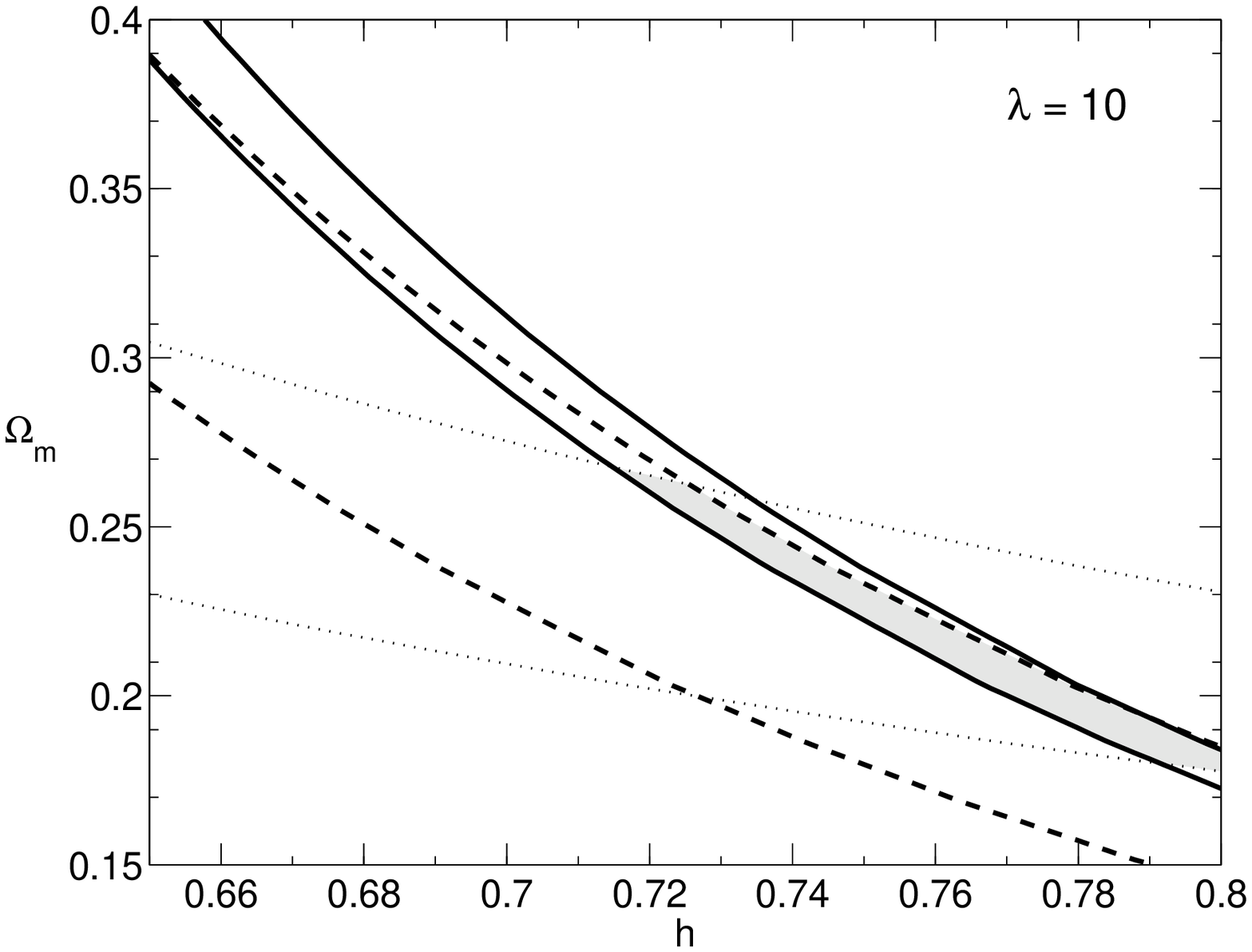}
 \includegraphics[height=5cm]{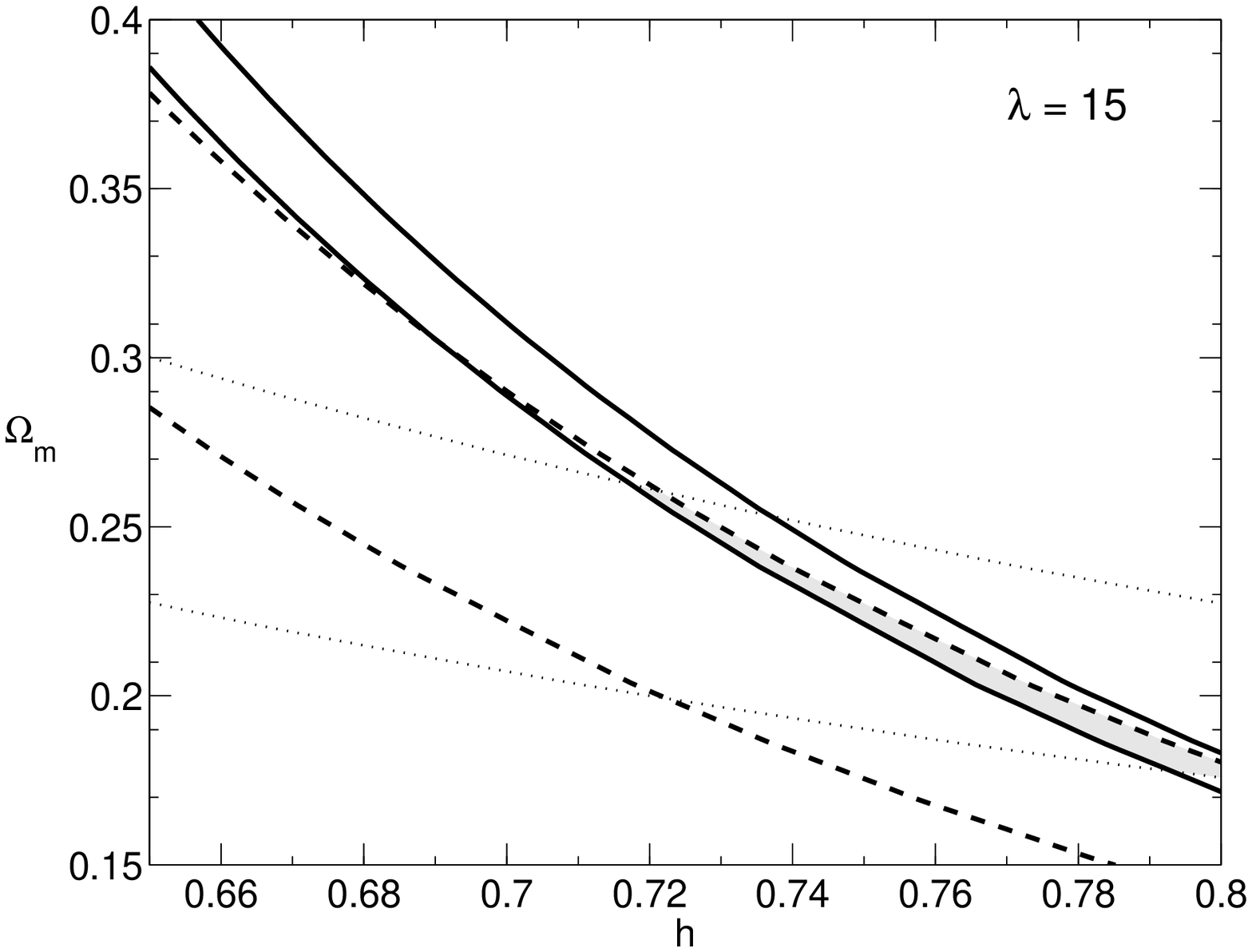}
 \includegraphics[height=5cm]{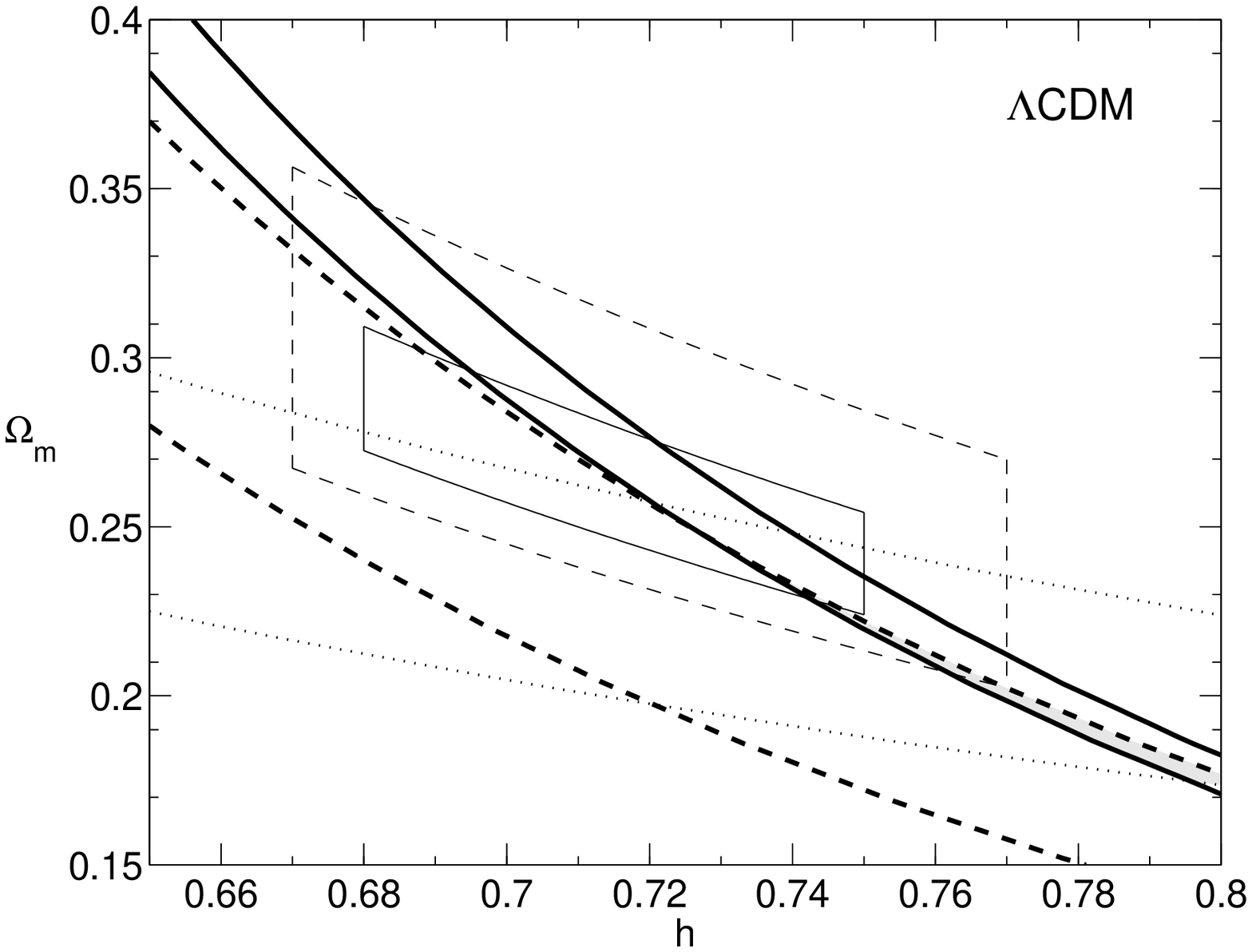}
  \caption{\label{fig:neq1} Contour plots of the first and
 second Doppler peaks location  in the
 $(\Omega_m, h)$ plane for the potential Eq.~(\ref{eq:pot}),
 with A=0.0025, for different values of $\lambda$ and spectral index $n_s=1$.
 Full and dashed contours correspond to WMAP's bounds on, respectively,
 $\ell_{p_1}$ and $\ell_{p_2}$, see Eq.~(\ref{eq:wmap}).
 The dotted contours correspond to $\sigma_8$ constraints.
 The intersection of all curves leads to the shaded allowed
 regions. We also show
 the corresponding contours for the $\Lambda$CDM model with $n_s=1$; the
 full and dashed 
 boxes  
indicate,   bounds from WMAP only  and a combination of WMAP and 
other
 experiments on  $h$ and $\Omega_m h^2$,  Eqs.~(\ref{eq:wmapb}) and 
 (\ref{eq:wmapc}) respectively.}
\end{center}
\end{figure*}

In an idealised model of the primeval plasma, there is a simple relation
between the location of the $m$-th peak and the acoustic scale, namely
$\ell_m\approx m \ell_A$. However, the location of the peaks is slightly 
shifted
by driving effects and
this can be compensated by  parameterising the location of the $m$-th
peak, $\ell_m$, as in \cite{Hu2001,Doran2}

\beq 
\ell_{p_m} \equiv \ell_A \left(m - \varphi_m\right)
\equiv \ell_A (m -\bar \varphi-\delta \varphi_m)~, 
\label{eq:lm}
\eeq
where $\bar \varphi\equiv \varphi_1$ is the overall peak shift and
 $\delta \varphi_m\equiv \varphi_m-\bar \varphi$ is the relative shift of the
 m-th
 peak relative to the first. Eq.~(\ref{eq:lm}) is correct also for
 the location of troughs if we set, for instance, 
 $m=3/2$  for the first trough and $m=5/2$ for the second trough.
Although is not in general possible to derive analytically
a relationship between the cosmological parameters and the peak shifts,
 one can use fitting formulae that describe their dependence on these
parameters. Doran and Lilley \cite{Doran2} give  accurate analytic
 approximations for the first three peaks and first trough, which can be
 found in the Appendix for convenience. Notice that,
 as the authors of
 Ref.~\cite{Doran2} point out, although these formulae
 were obtained using a standard exponential potential, one expects the results
 to be fairly independent of the form of the potential (since  the shifts
 are almost independent of post recombination physics) unless it is
 qualitatively
 very different from the exponential potential before last scattering.

\begin{figure*}[ht!]
\begin{center}
 \includegraphics[height=5cm]{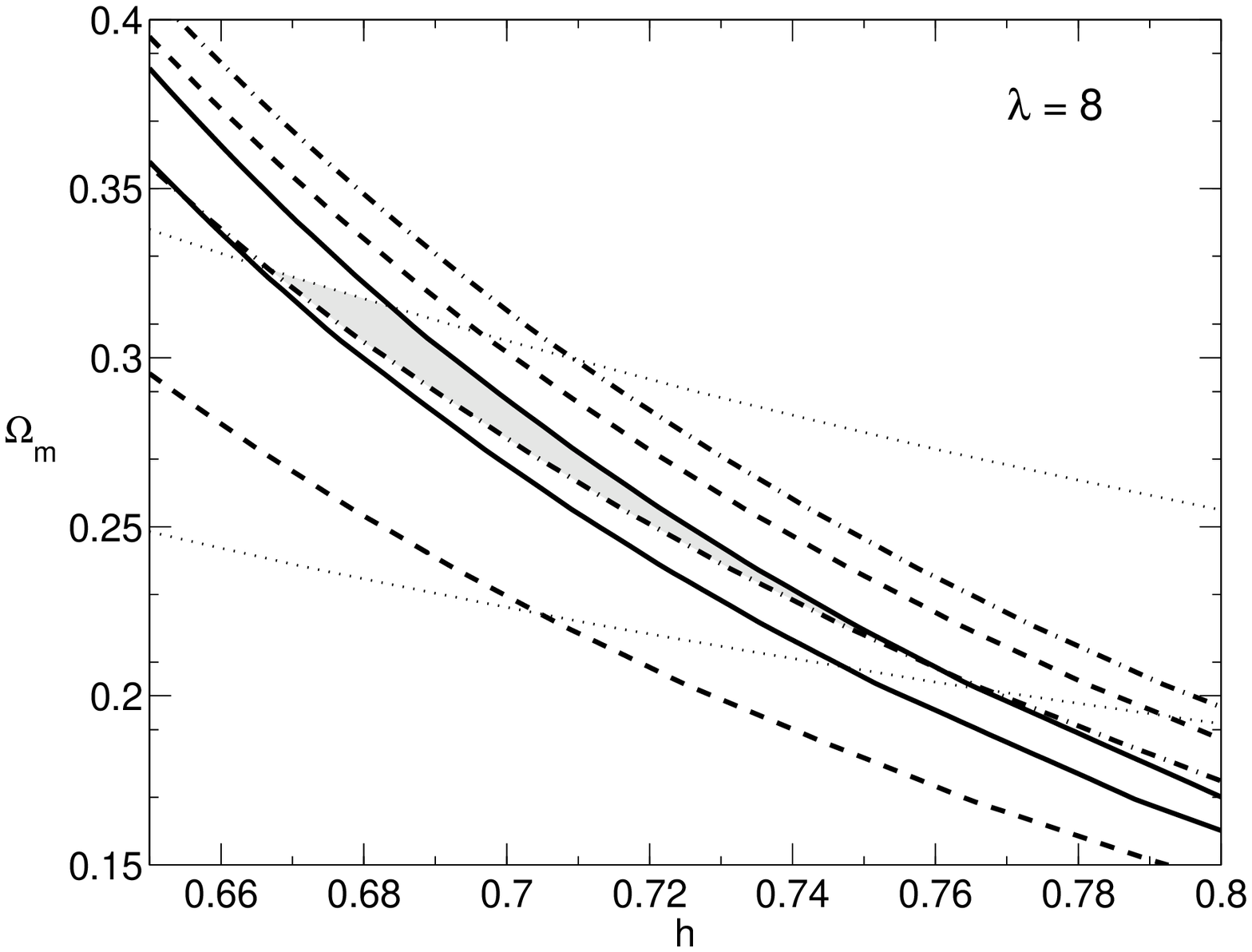}
 \includegraphics[height=5cm]{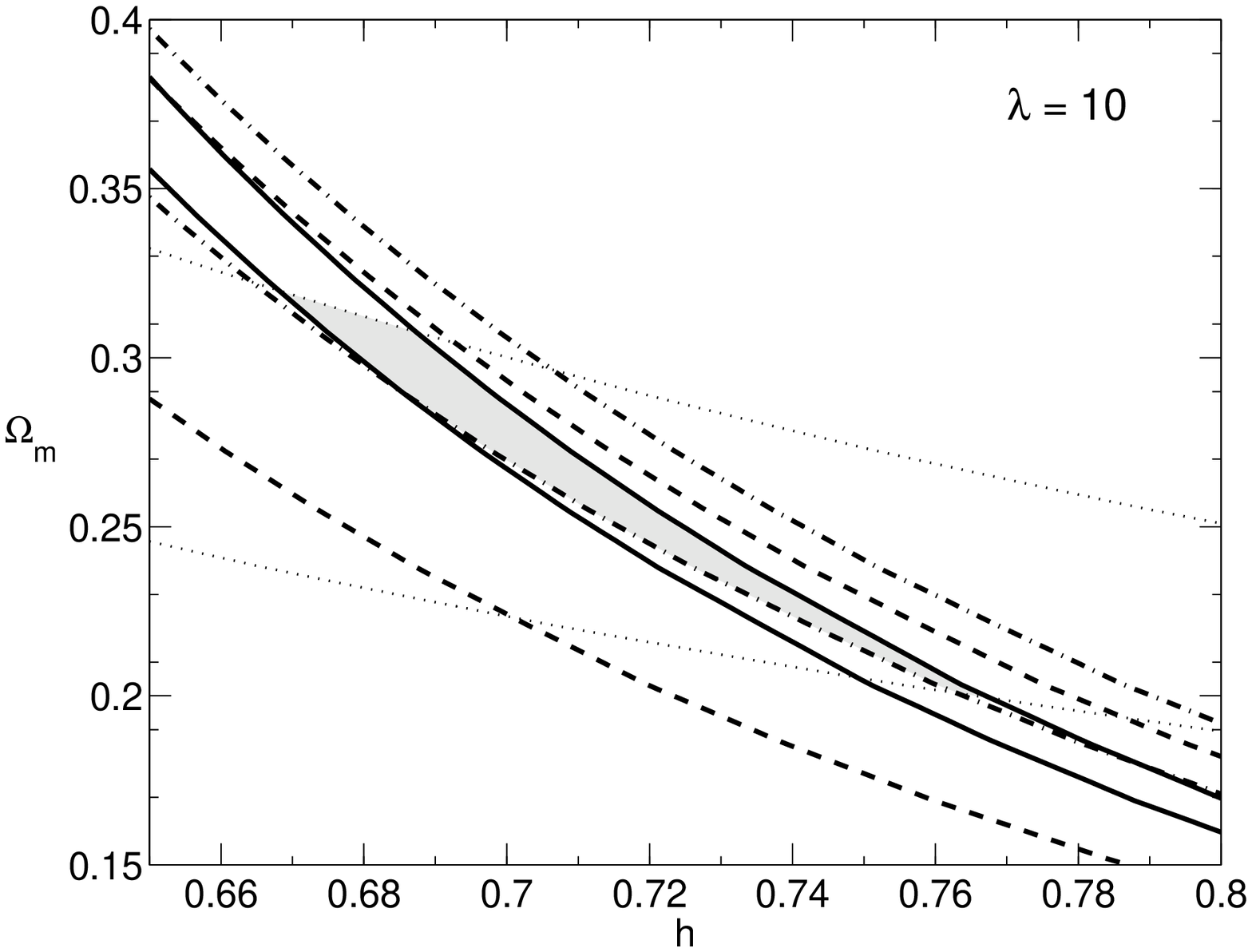}
 \includegraphics[height=5cm]{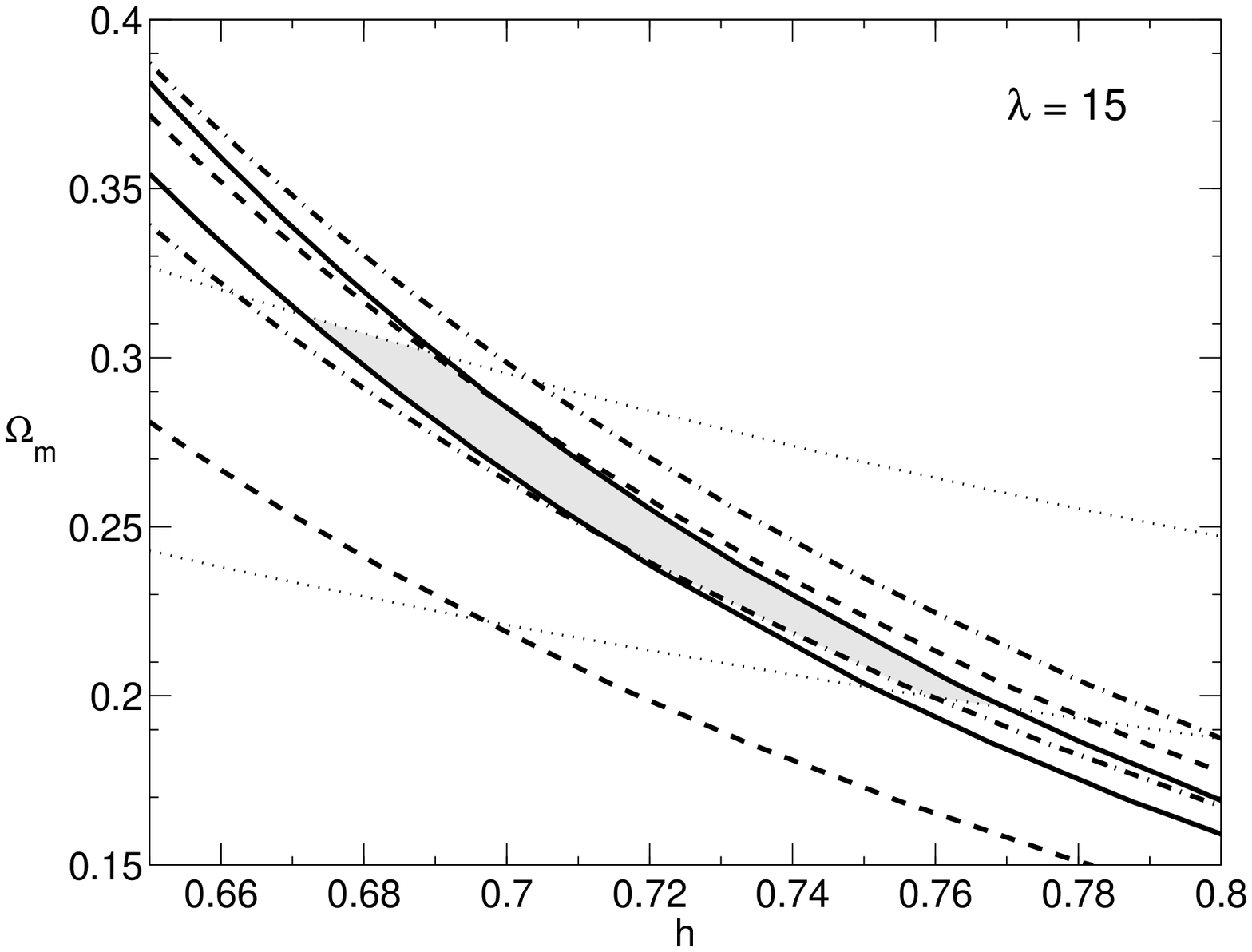}
 \includegraphics[height=5cm]{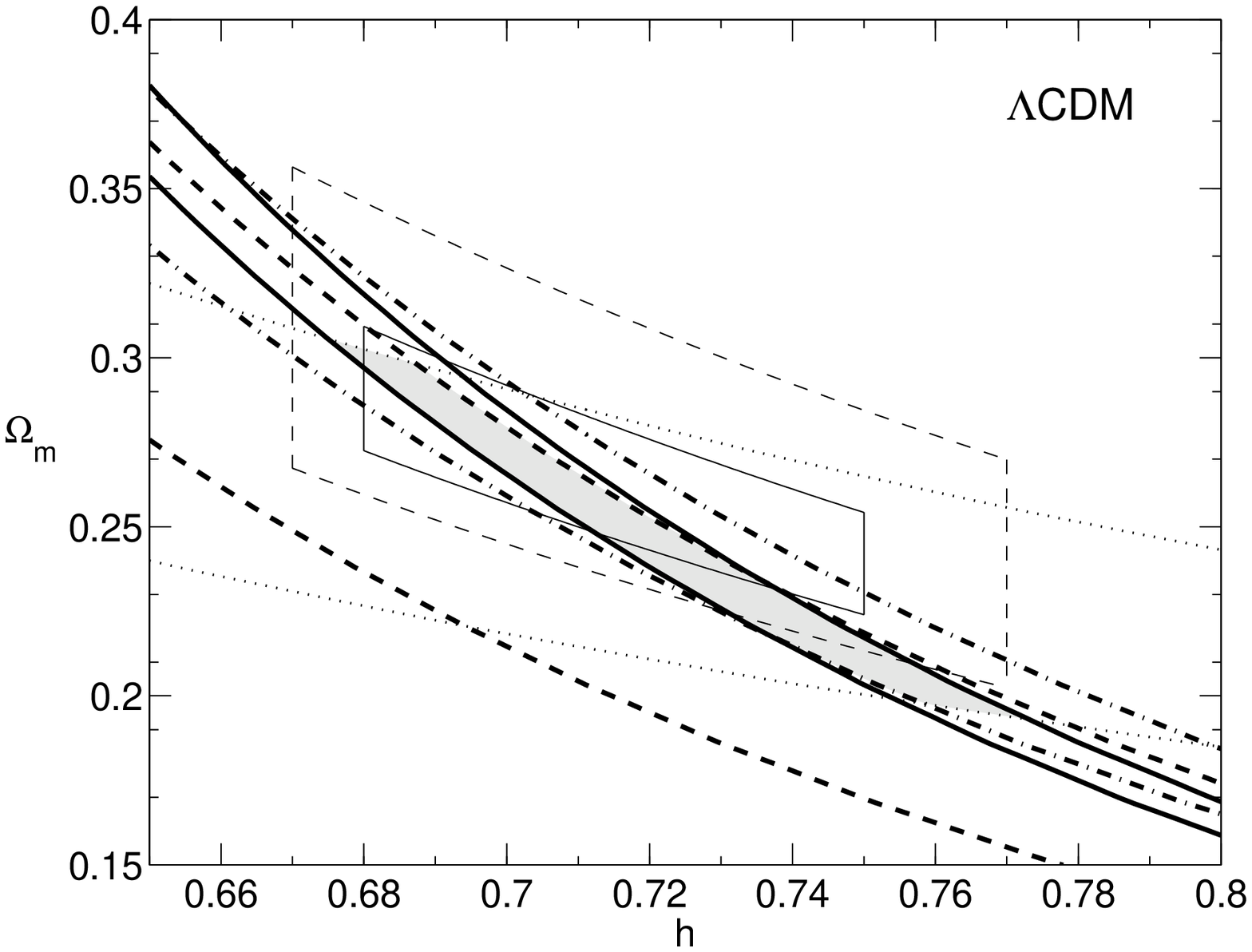}
  \caption{\label{fig:neq95} As for Fig.~1 but with $n_s=0.95$ and the contours
 corresponding to the observational bounds on the first trough,
 Eq.~(\ref{eq:wmap}), are also shown (dot-dashed contours).  }
\end{center}
\end{figure*}

\begin{figure*}[ht!]
\begin{center}
 \includegraphics[height=5cm]{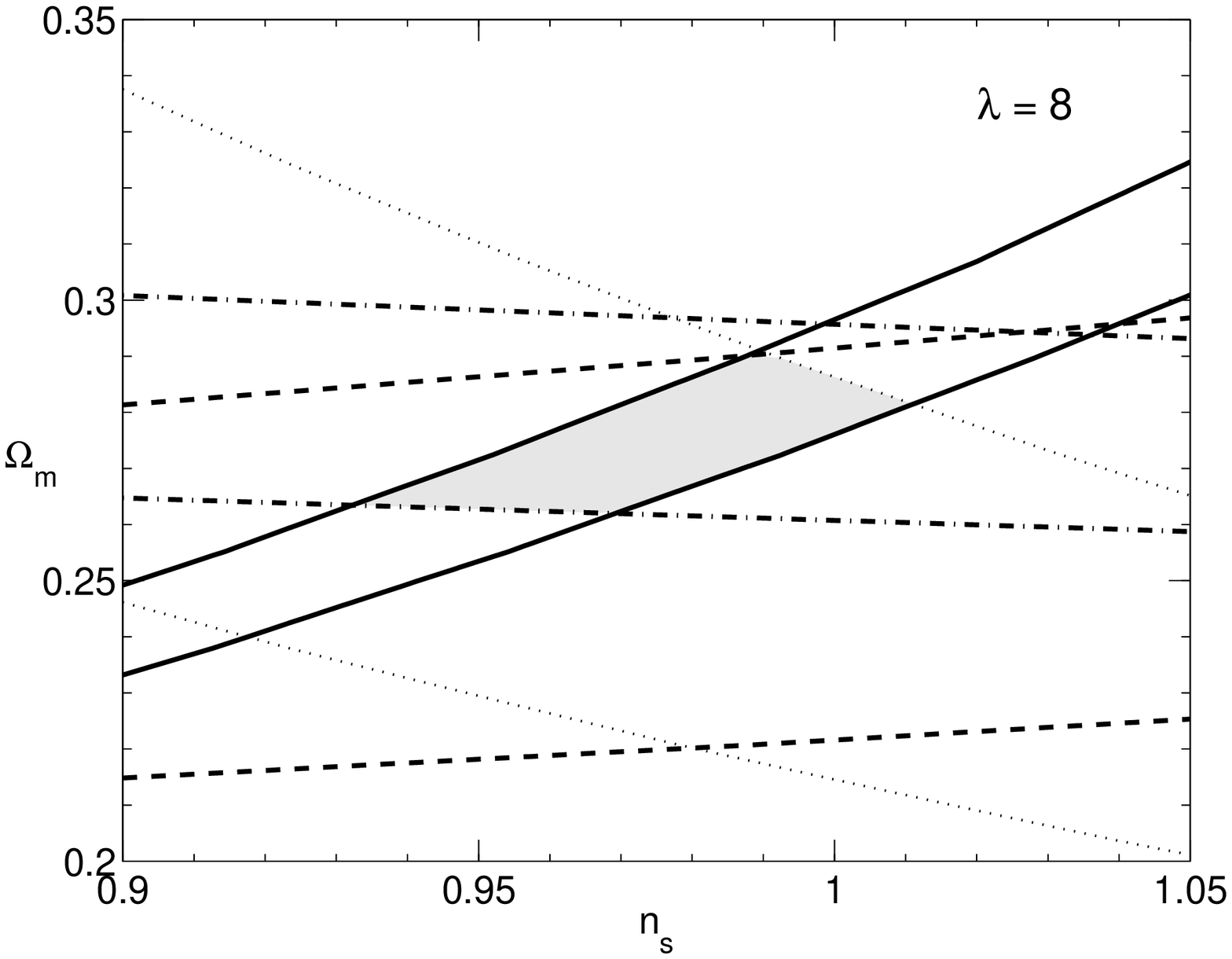}
 \includegraphics[height=5cm]{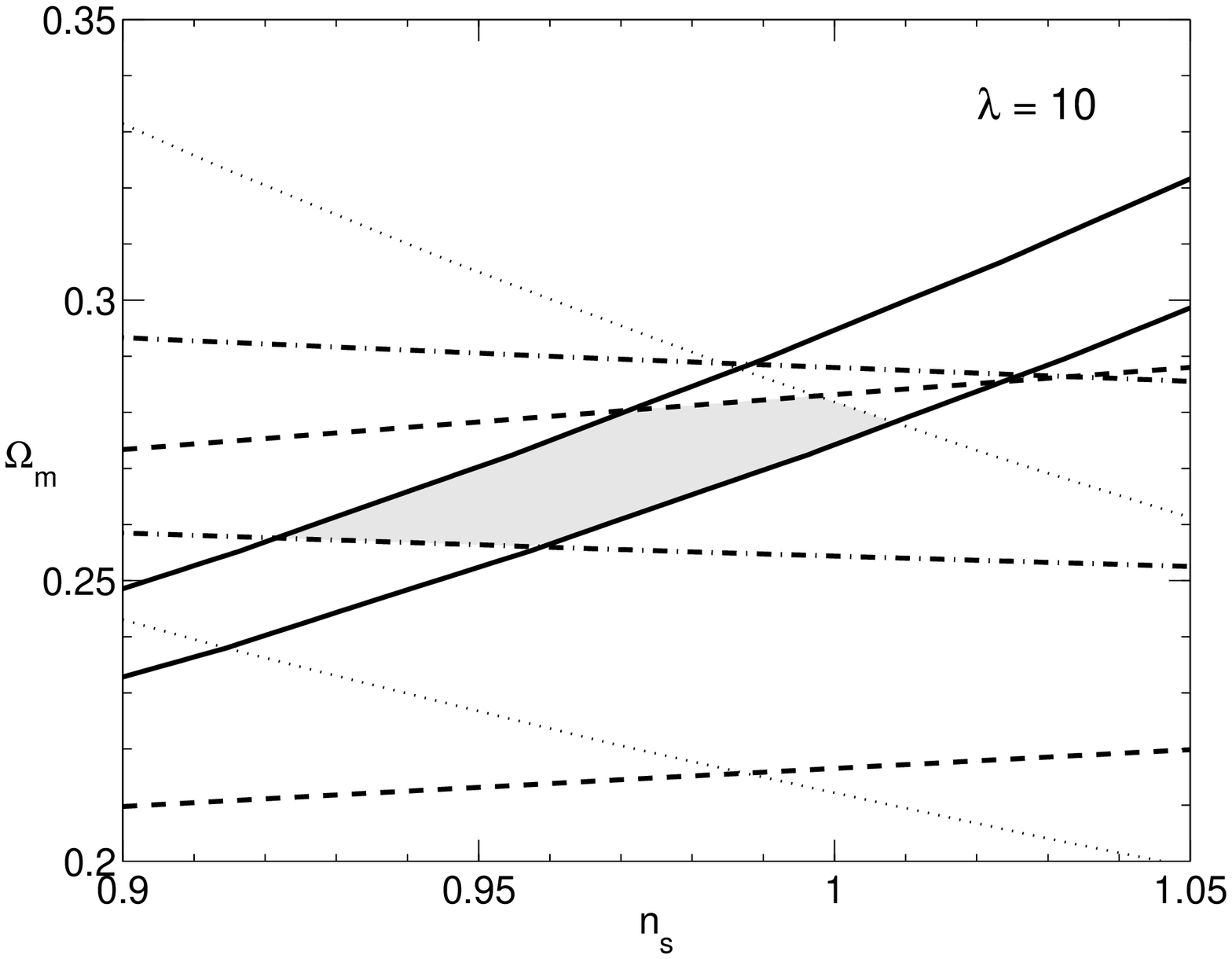}
 \includegraphics[height=5cm]{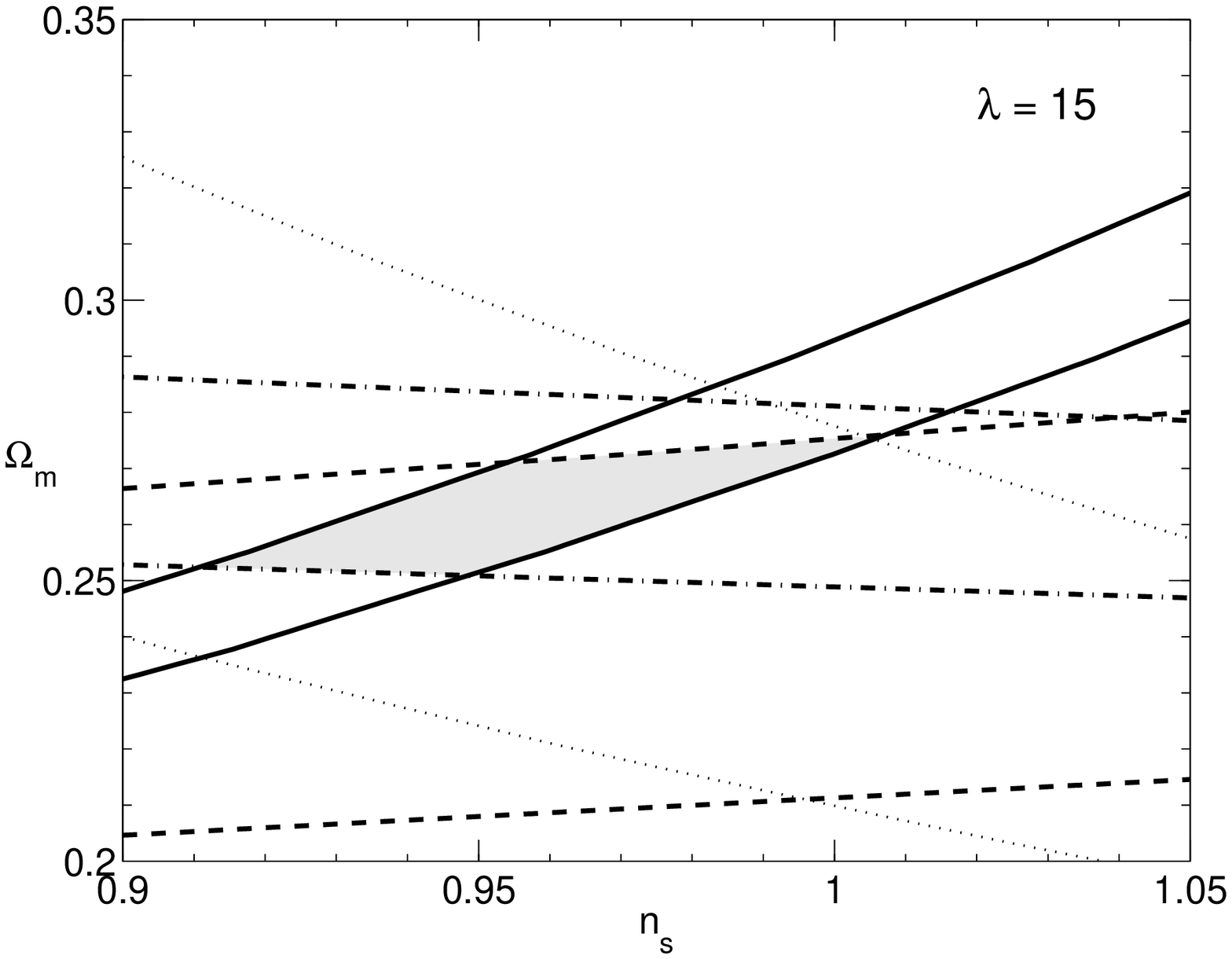}
 \includegraphics[height=5cm]{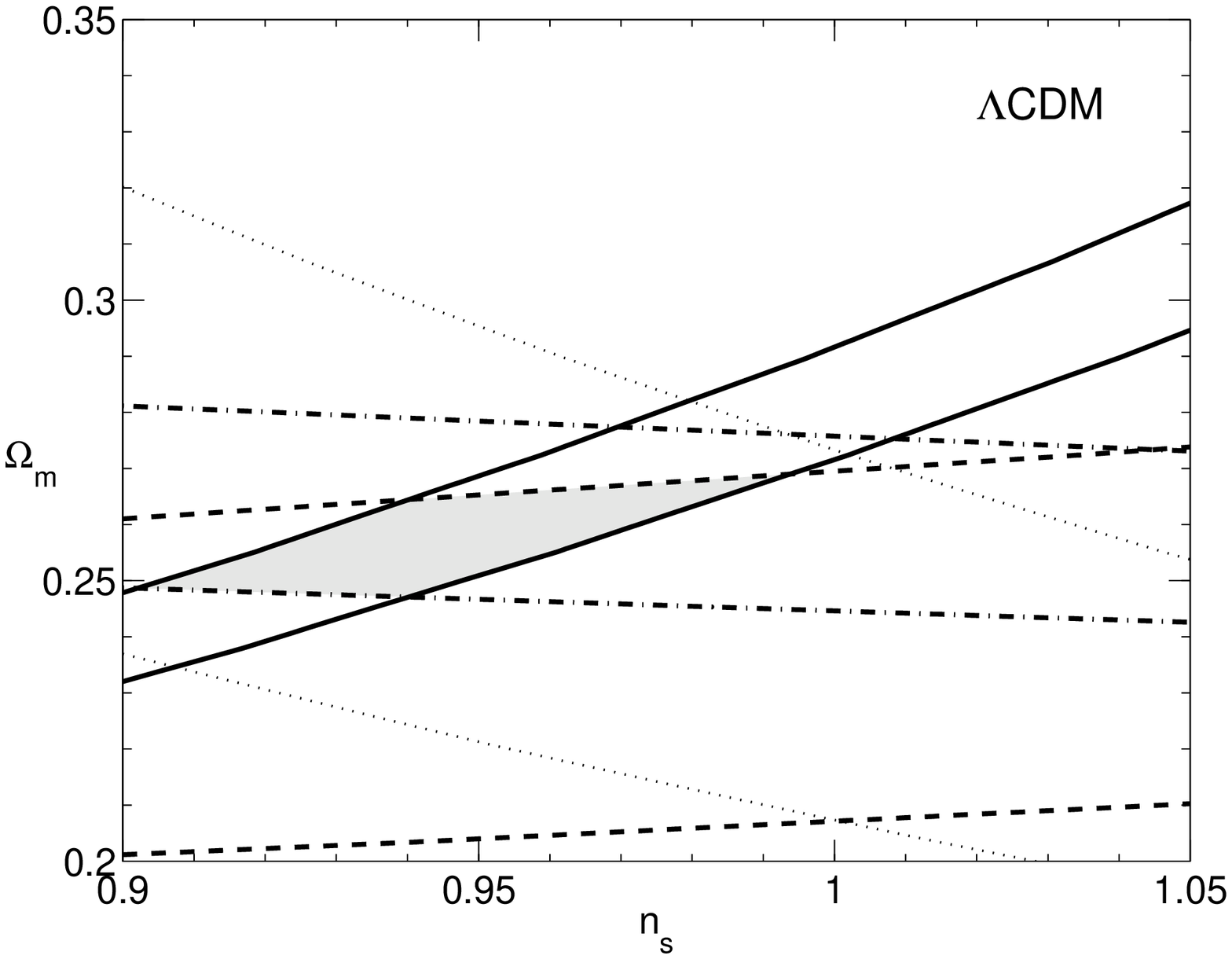}
  \caption{\label{fig:h71} Contour plots of the locations of the 
first and second
  peaks and first trough  in the
 $(\Omega_m, n)$ plane for the potential Eq.~(\ref{eq:pot}),
 with A=0.0025, for different values of $\lambda$ and  $h=0.71$.
 Full, dashed and dot-dashed contours correspond to WMAP's bounds on
 $\ell_{p_1}$,  $\ell_{p_2}$  and $\ell_{d_1}$,  Eq.~(\ref{eq:wmap}),
 respectively.
The dotted contours correspond to $\sigma_8$ constraints.
 We also show  the equivalent  plot for the $\Lambda$CDM model
 with $h=0.71$. }
\end{center}
\end{figure*}

The locations of the first two acoustic peaks and the first trough,
from the WMAP measurements of 
the CMB temperature angular power spectrum are \cite{MAP1}

\beqa
\ell_{p_1} &=& 220.1\pm 0.8~,\nonumber\\  
\ell_{p_2} &=& 546\pm 10~,\nonumber\\  
\ell_{d_1} &=& 411.7\pm 3.5~;   
 \label{eq:wmap}
\eeqa
notice that all uncertainties are 1$\sigma$ and include calibration and
 beam errors.
The location of the third peak is given by BOOMERanG measurements \cite{BOOM1}

\beq
\ell_{p_3} = 825^{+10}_{-13}~.
\label{eq:l3}
\eeq

We have studied the location of the first three peaks and first trough 
in the $(\Omega_m,\ h,\ n_s)$ cosmological parameter space 
  for the potential Eq.~(\ref{eq:pot}),
 for different values of $A$ and $\lambda$,
 slicing through the  $(\Omega_m, h)$ an  $(\Omega_m, n_s)$ planes. For
 each value
 of $\lambda$ and A, $\phi_0$ is 
chosen such that $\Omega_{tot}=1$. Given the rather strict bound
 $\Omega_b h^2=0.0224\pm0.0009 $ \cite{MAP1}, 
 we assume
 hereafter $\Omega_b h^2=0.0224$.
 
Our analysis in the $(\Omega_m, h)$ plane shows that, for $n_s=1$,
 the strongest constraints
 arise from the positions of the first and second peaks and, for clarity,
 we plot only the $\ell_{p_1}, \ell_{p_2}$ contours corresponding to the WMAP
 bounds on these quantities,  Eq.~(\ref{eq:wmap}), for different values of
 $\lambda$ and $A=0.0025$ (see Fig.~\ref{fig:neq1}); also shown are  the 
 $\ell_{p_1}, \ell_{p_2}$ 
contours corresponding
 to the $\Lambda$CDM  model for the same values of the cosmological parameters,
 where the  dashed box represents  
WMAP's bounds on 
$h$ and $\Omega_m h^2$, namely \cite{MAP1}

\beq
\Omega_m h^2=0.14\pm 0.02 ,~~~h=0.72\pm 0.05
\label{eq:wmapb}
\eeq
The full box corresponds to the bounds obtained on these quantities
 from a combination of WMAP data with other CMB experiments (ACBAR and CBI),
 2dFGRS measurements and Lyman $\alpha$ forest  data  \cite{MAP1}

\beq
\Omega_m h^2=0.135^{+0.008}_{-0.009} ,~~~h=0.71^{+0.04}_{-0.03}~.
\label{eq:wmapc}
\eeq
For $n_s=0.95$, the strongest constraints come 
from WMAP's bounds on  $\ell_{p_1},
 \ell_{p_2}$ and  $\ell_{d_1}$, see Fig.~\ref{fig:neq95}.

 Finally, the results of our analysis in the $(\Omega_m, n_s)$
  plane,
 with $h=0.71$, are shown in Fig.~\ref{fig:h71}, where we plot
 the same contours as for  Fig.~\ref{fig:neq95}.

\section{Constraints from $\sigma_8$}

We have also studied  constraints resulting from  $\sigma_8$, the
 {\it rms} density 
fluctuations averaged over $8 h^{-1} \textrm{Mpc}$ spheres.

Ref.~\cite{Doran4} gives an estimate of the CMB-normalized $\sigma_8$-value
 for 
a very general class of 
quintessence models  from 
 $\Omega_\phi(a),\ w_\phi(a)$ 
and the $\sigma_8$-value of the $\Lambda$CDM model with 
the same amount of dark energy today
 $\Omega_\Lambda^0=\Omega_\phi^0(\Lambda)$:

\beq
 \label{eq:main}
 {\sigma _8^\phi \over \sigma _8^\Lambda}\approx
\left( a_{eq}\right)^{ 3\, {\bar\Omega}_\phi^{sf}/ 5}
 \left(1-\Omega_{\Lambda}^0 \right)^{-\left (1+ \bar w ^{-1}\right)/5}
 \sqrt{\frac{\tau _0^\phi}{\tau _0^\Lambda}}.
\eeq
where $a_{eq}$ is the scale factor at matter-radiation equality,
 $ {\bar\Omega}_\phi^{sf}$
 is an average value 
for the fraction of dark energy during structure formation {\it i.e.}
during the matter dominated era, before $\Omega_\phi$ starts growing
 rapidly at 
scale factor $a_{tr}$

\beq
 {\bar\Omega}_\phi^{sf} \equiv [ \ln{a_{tr}}-\ln{a_{eq}} ]^{-1}
 \int_{\ln a_{eq}}^{\ln a_{tr}} \Omega_\phi(a)\  d \ln a~.
\eeq
The 
effective equation of state
 $\bar{w}$ is an
average value for $w_\phi$ during the time in which $\Omega_\phi$ is growing 
rapidly:

\beq
{\bar{w}}^{-1}=\frac{\int _{\ln a_{\rm tr}}^0 \Omega_\phi(a)w^{-1}(a)
                   d \ln{a}}
                   {\int _{\ln a_{\rm tr}}^0 \Omega_\phi(a) d  \ln{a}}.
\eeq

In order to compute $\sigma_8$ for the $\Lambda$CDM model, we use the
 definition

\beq
\sigma_8^2 \equiv \int_0^\infty {dk\over k}\ \Delta^2(k)\,
 \left( {3j_1(kr)\over kr}\right)^2,
\label{eq:sigma8}
\eeq
with $r=8\ h^{-1}$ Mpc and 

\beq
\Delta^2(k) = \delta_H^2 \left( {k\over H_0}\right)^{3+n} T^2(k),
\label{eq:delta}
\eeq
where  $T(k)$
 is the transfer function describing the processing of the initial
fluctuations, for which we use the fitting function \cite{Bardeen}

\beqa
T(q) & = & \frac{\ln \left(1+2.34q \right)}{2.34q} \times
        \\ \nonumber
& & \left[1+3.89q+(16.1q)^2+ \right. \\ \nonumber
& & \left. (5.46q)^3+(6.71q)^4\right]^{-1/4}~,
\label{eq:Trans}
\eeqa
with  \cite{Sugiyama}

\beqa
q &=& {k\over h\Gamma}\mbox{Mpc}~,\nonumber\\
\Gamma &=& \Omega_m h\exp\left[-\Omega_b\left(1+{\sqrt{2h}\over
 \Omega_m}\right)\right]
\label{eq:q}
\eeqa
and $\delta_H$ is the density perturbation at horizon crossing. A fit to the
 four-year COBE data gives  ~\cite{Bunn}

\beq
10^5\delta_H=1.94\Omega_m^{-0.785-.05\ln\Omega_m}\exp[a\tilde
 n+b\tilde n^2]
\label{eq:delH}
\eeq
where $\tilde n=n_s-1,\ a=-0.95$ and $b=-0.169$ (assuming there are 
no gravitational waves).
Taking into account reionization effects, Eq.~(\ref{eq:delH}) should be 
corrected;
we use the fitting formula of
 Ref.~\cite{Louise}

\beq
{\delta_H(\tau)\over \delta_H(\tau=0)}=1+0.76 \tau-1.96\tau^2+1.46\tau^3~,
\label{eq:delHtau}
\eeq
where $\tau$ is the optical depth. This formula 
 is reliable up to $\tau\approx 0.5$; we use $\tau=0.11$, which is
 within the range of   WMAP's bound,
 $\tau=0.166^{+0.076}_{-0.071}$ \cite{MAP1}.

We compare our results for $\sigma_8$ with large scale
 structure data. The recent study of the mass function of 300
 clusters at redshifts $0.1<z<0.2$ using  early SDSS data yields
 $\sigma_8\Omega_m^{0.6}=0.33 \pm 0.03$  \cite{Bahcall}. Other cluster 
analysis (using eg. BCS, REFLEX and ROSAT data) yield different values:
 $\sigma_8=0.43\Omega_m^{-0.38}$
 \cite{Reiprich} ,
 $\sigma_8=(0.508\pm 0.019)\Omega_m^{-(0.253\pm 0.024)}$ \cite{Allen}, 
$\sigma_8=(0.7\pm 0.06)(\Omega_m/0.35)^{-0.44}(\Gamma/0.2)^{0.08}$ 
\cite{Seljak} and $\sigma_8=0.38\Omega_m^{-0.48+0.27 \Omega_m}$ 
\cite{Viana}.

The results of our study of $\sigma_8$ constraints for the model of
 Eq.~(\ref{eq:pot}) are shown in Figs. 1 $-$ 3 (dotted curves), where we have
 plotted the largest $\sigma_8$ contours that are compatible both 
with the large scale
 structure fits mentioned above and the  CMB computation,
 Eqs.~(\ref{eq:main}) - (\ref{eq:delHtau}).

\section{Discussion and conclusions}

Fig.~\ref{fig:neq1}  shows that, as $\lambda$
 increases, the allowed region
 becomes smaller and more similar to the
 $\Lambda$CDM model results; in particular, for $\lambda=15$, the allowed
 region is only marginally  larger than the one for $\Lambda$CDM model,
 which is quite small.

Notice that a lower bound on $\lambda$ already exists from  standard 
Big Bang Nucleosynthesis (BBN) and the observed abundances of 
primordial nuclides, which implies $\Omega_\phi(\mev)<0.045$,
 or, considering a possible underestimation of systematic errors,
 the more conservative
 result
 $\Omega_\phi(\mev)<0.09$ 
 \cite{Bean}; these bounds require, respectively,
  $\lambda>9$ and  $\lambda > 6.5$, for 
 the  model we are considering.

 For $n_s=0.95$,  Fig.~\ref{fig:neq95} shows that 
 the allowed
 region becomes larger until $\lambda\sim 15$ but does not change 
significantly for $10\lsim \lambda \lsim 18$. Again, we find that, as $\lambda$
 increases, $\lambda \gsim 18$, the model's results  become more
 similar to the ones for the
$\Lambda$CDM model.

Regarding the  results of our analysis in the $(\Omega_m, n_s)$  plane,
 with $h=0.71$,  Fig.~\ref{fig:h71},  we see 
that the allowed
region becomes  smaller and shifts towards smaller values of $n_s$ and
 $\Omega_m$ as  $\lambda$ increases, in which
 case there is again a similarity with the $\Lambda$CDM model results.

We have also studied the dependence of the first three peaks  and
 first trough locations on parameter A. We find that this dependence is
 extremely small and can safely be neglected. We notice that the values of $A$
 and $\lambda$ we have previously considered all correspond to the permanent
 acceleration regime; however, as  should be expected,  changing $A$ so as to
reach  the transient acceleration regime does not  alter  the
analysis  since
  this regime is not   significantly different
 from the one where acceleration is permanent  until the present,
 and therefore  peak positions should not be affected.

Hence, the model's behaviour depends essentially on parameter $\lambda$,
 which measures the amount of ``early quintessence'' (using the terminology
 of Ref. \cite{Caldwell2003})
 since ${\bar\Omega}_\phi^{ls}\sim 3/\lambda^2$, where
   ${\bar\Omega}_\phi^{ls}$ is the average fraction of dark
 energy before last scattering

\beq
{\bar \Omega}_{\phi}^{ls}=\tau_{ls}^{-1} \int_0^{\tau_{ls}} \Omega_\phi(\tau)
 d\tau~.
\label{eq:baro}
\eeq
In fact, we obtain, for $\lambda=8$,  
 $0.039<{\bar \Omega}_\phi^{ls}<0.043$, for the cosmological parameter range
 we are 
considering.

 We conclude that, for $n_s\approx 1$, the $\Lambda$CDM model becomes 
increasingly disfavoured compared with the quintessence model of
 Eq.~(\ref{eq:pot}) as the amount of early quintessence becomes higher
 ($\lambda\lsim 15$). 
For $n_s < 1$, the opposite is true {\it i.e} $\Lambda$CDM is favoured as 
compared to quintessence if
 $\lambda\lsim 15$. 
Notice that, as $\lambda$ increases ($\lambda\gsim 18$), independently of the
 value of $n_s$, the
 model's results become 
comparable to $\Lambda$CDM's, as should be expected since 
 ${\bar \Omega}_\phi^{ls}$ decreases. 
Moreover, we see  that quintessence is distinguishible from $\Lambda$CDM only
 for $h<0.73$ and $n_s\approx 1$.

Finally, we would like to mention that
 the non-negligible values we obtain for 
 ${\bar\Omega}_\phi^{sf}$  (e.g. for
 $\lambda=8$, we get  $0.022<{\bar \Omega}_\phi^{sf}<0.026$),
 typical of early quintessence
 models,  will lead,  as shown in Ref.~ \cite{Caldwell2003}, to suppressed
 clustering
 power on small length scales 
 as suggested by WMAP/CMB/large scale structure
 combined data.

\begin{acknowledgments}

\noindent
The work of T. Barreiro is fully financed by the  Funda\c c\~ao para a 
Ci\^encia e a Tecnologia (FCT) grant PPD/3512/2000.
 M.C. Bento 
 acknowledges the partial support of FCT
under the grant POCTI/1999/FIS/36285. The work of A.A. Sen is fully 
financed by the same grant. N.M.C. Santos is supported by FCT grant
SFRH/BD/4797/2001.

\end{acknowledgments}
\appendix*
\section{}
We have used the analytic approximations for the phase shifts found in
 Ref.~\cite{Doran2}, which we reproduce here for completeness.
 The overall phase shift is given by
\beq
{\bar\varphi}=(1.466-0.466 n_s)\left[a_1 r_*^{a_2}
+ 0.291 {\bar\Omega}_{\phi}^{ls} \right]~,
\label{eq:phibar}
\eeq   
where

\beqa
a_1 &=& 0.286+0.626\omega_b\nonumber\\
a_2 &=& 0.1786-6.308\omega_b+174.9\omega_b^2-1168\omega_b^3
\label{eq:as}
\eeqa
with $\omega_b=\Omega_b h^2$, are fitting coefficients, 
${\bar \Omega}_{\phi}^{ls}$ is given by Eq.~(\ref{eq:baro})
and

\beq
r_*\equiv \rho_{r}(z_{ls})/\rho_{m}(z_{ls})
\eeq
is the ratio of radiation to matter  at
 decoupling. A convenient fitting formula for $z_{ls}$
 is \cite{Hu1996}

\beq
z_{ls}=1048[1+0.00124 w_b^{-0.738}[1+g_1 w_m^{g_2}]~,
\label{eq:zls}
\eeq
where

\beqa
g_1 &=& 0.0783 w_b^{-0.238} [1+39.5 w_b^{0.763}]^{-1}~,\nonumber\\
g_2 &=& 0.56 [1+21.1 w_b^{1.81}]^{-1}~.
\label{eq:coef}
\eeqa
\

The relative shift of the first acoustic peak is zero, $\delta\varphi_1=0$,
and the relative shifts of the  second and third peaks are given by

\beq
\delta\varphi_2=c_0-c_1r_*-c_2r_*^{-c_3}+0.05(n_s-1)~,
\label{eq:delphi2}
\eeq
with
\beqa
c_0 &=& -0.1+\left(0.213-0.123{\bar\Omega}_{ls}^\phi \right)\nonumber\\
        &&\times\exp\left\{-\left(52-63.6 {\bar\Omega}_{ls}^\phi \right)
          \omega_b\right\},\nonumber\\
c_1 &=& 0.015+0.063\exp\left(-3500\omega_b^2\right), \nonumber\\
c_2 &=& 6\times 10^{-6}+0.137(\omega_b-0.07)^2,\nonumber\\
c_3 &=& 0.8+ 2.3 {\bar\Omega}_{ls}^\phi
+\left(70-126{\bar\Omega}_{ls}^\phi\right)\omega_b~,
\label{eq:cs}
\eeqa
and 

\beq
\delta\varphi_3=10-d_1r_*^{d_2}+0.08(n_s-1)~,
\eeq
with
\beqa
d_1 &=& 9.97+\left(3.3-3{\bar\Omega}_{ls}^\phi\right)\omega_b,\nonumber\\
d_2 &=& 0.0016-0.0067{\bar\Omega}_{ls}^\phi +
\left(0.196-0.22{\bar\Omega}_{ls}^\phi\right)\omega_b\nonumber\\
&& + \left(2.25+2.77{\bar\Omega}_{ls}^\phi\right)\times  10^{-5}\omega_b^{-1}.
\label{eq:ds}
\eeqa

The relative shift of the first trough is 

\beq
\delta\varphi_{3/2}=b_0+b_1r_*^{1/3}\exp(b_2r_*)+0.158(n_s-1)~
\label{eq:delphi}
\eeq
with 

\beqa
b_0 &=& -0.086-0.079 {\bar\Omega}_{\phi}^{ls}
       -\left( 2.22-18.1{\bar\Omega}^{ls}_\phi \right) \omega_b\nonumber\\
    &&-\left(140+403{\bar\Omega}^{ls}_\phi \right)\omega_b^2~,\nonumber\\
b_1 &=& 0.39-0.98{\bar\Omega}^{ls}_\phi
       -\left(18.1-29.2{\bar\Omega}_{ls}^\phi\right)\omega_b\nonumber\\
    && +440\omega_b^2,\\
b_2 &=& -0.57-3.8\exp({-2365\omega_b^2})~,
\label{eq:bs}
\eeqa

Notice that the deviation of the acoustic extrema locations calculated
 using the above formulae
 from the values obtained by CMBfast code is 
$< 3\%$ for a sufficiently wide range of parameters.



\begin{thebibliography}{99}



\bibitem{Ratra} B. Ratra, P.J.E.  Peebles,  \PR {\bf  D37} 3406 (1988);
 \AJL {\bf 325} 117 (1988).
 
\bibitem{Wetterich} C. Wetterich, \NP {\bf B302} 668 (1988). 

\bibitem{Caldwell} R.R.  Caldwell, R.  Dave, P.J.  Steinhardt, \PRL
{\bf 80}, 1582 (1998).
 
\bibitem{Ferreira} P.G.  Ferreira, M.  Joyce, \PR {\bf  D58},023503  (1998)
.

\bibitem{Zlatev} I. Zlatev, L. Wang, P.J. Steinhardt, \PRL {\bf 82}, 986
(1999).

\bibitem{Binetruy} P. Bin\'etruy, \PR {\bf D60}, 063502 (1999) .

\bibitem{Kim} J.E. Kim, \JHEP {\bf}9905,  022 (1999).

\bibitem{Barreiro} T.~Barreiro, E.~J.~Copeland, N.~J.~Nunes, Phys. Rev. D
 {\bf 61}, 127301 (2000).

\bibitem{Bento} M.C. Bento and  O. Bertolami,  \GRG {\bf  31}, 1461 (1999);
 M.C. Bento, O. Bertolami, P.T. Silva, \PL {\bf B498}, 62  (2001).

\bibitem{Uzan} J.P. Uzan, \PR {\bf D59}, 123510 (1999); 
T. Chiba, \PR {\bf D60} 083508  (1999);
L. Amendola, \PR {\bf D60}, 043501 (1999);
O. Bertolami, P.J. Martins, \PR {\bf D61}, 064007 (2000);
N. Banerjee, D. Pav\'on, \PR {\bf D63}, 043504 (2001) ; \CQG {\bf 18}  593
(2001);
A.A. Sen, S. Sen, S. Sethi, \PR {\bf D63}, 107501   (2001);
A.A. Sen, S. Sen, \MPL {\bf A16},  1303 (2001).

\bibitem{Fujii} Y. Fujii, \PR {\bf D61},  023504 (2000).

\bibitem{Masiero} A. Masiero, M. Pietroni and F. Rosati, 
\PR {\bf D61}, 023504 (2000) .

\bibitem{Bento1} M.C. Bento, O. Bertolami, N. C. Santos, \PR {\bf D65}, 067301 
(2002) 
.

\bibitem{Chap} A. Kamenshchik, U. Moschella,
 V. Pasquier, \PL {\bf 511}, 265 (2001);
 M.C. Bento, O. Bertolami and A.A. Sen, \PR {\bf D66}, 043507 (2002);
  N. Bili\'c, G.B. Tupper, R.D. Viollier, \PL {\bf B535}, 17 (2002).

\bibitem{card} K. Freese, M. Lewis, Phys. Lett. {\bf B540},1 (2002);
 S. Sen, A.A. Sen, astro-ph/0211634. 

\bibitem{Doran1} M. Doran, M. Lilley, J. Schwindt, C. Wetterich, \AJ {\bf 559},
 501 (2001).


\bibitem{BOOM}
P.~de~Bernardis {\it et al.}, \nat {\bf 404}, 955 (2000);
A.E.~Lange {\it et al.},Phys. Rev. {\bf D63}, 042001 (2001);
C.B. Netterfield {\it et al.}, Astrophys. J.  {\bf 571}, 604 (2002).

\bibitem{MAX}
S.~Hanany {\it et al.}, Astrophys.\ J.\  {\bf 545}, L5 (2000);
A.~Balbi {\it et al.}, Astrophys.\ J.\  {\bf 545}, L1 (2000)
[Erratum-ibid.\  {\bf 558}, L145 (2001)].

\bibitem{ARCH} A. Beno\^{\i}t {\it et al.}, { astro-ph/0210306}.  

\bibitem{MAP1} D.N. Spergel {\it et al}, { astro-ph/0302209}.
  

\bibitem{Hu} W. Hu, {astro-ph/0210696}.

\bibitem{Chap1}  M.C. Bento, O. Bertolami, A.A. Sen, astro-ph/0210468. 

\bibitem{Domenico} D. Di Domenico, C. Rubano, P. Scudellaro,
 {astro-ph/0209357}.

\bibitem{Doran3} M. Doran, M. Lilley, C. Wetterich, Phys. Lett. {\bf B528},
 175 (2002).

\bibitem{Brax} P. Brax, J. Martin, A. Riazuelo, Phys. Rev. {\bf D62},
 103505 (2000).

\bibitem{Tomo} M.~Kawasaki, T.~Moroi, T.~Takahashi, Phys. Rev. D {\bf 64},
 083009 (2001).

\bibitem{AS1} A. Albrecht, C. Skordis, \PRL 84, 2076 (2000).

\bibitem{AS2} A. Albrecht, C. Skordis, Phys. Rev. {\bf D66}, 043523 (2002).

\bibitem{Barrow} J. Barrow, R. Bean, J. Magueijo, 
 Mon. Not. Roy. Astron. Soc. {\bf 316}, L41 (2000). 

\bibitem{Hellerman} S. Hellerman, N. Kaloper, L. Susskind, JHEP 
{\bf 0106}, 003 (2001); 
W. Fischler, A. Kashani-Poor, R. McNess, S. Paban, JHEP {\bf 0107},
 003 (2001); 
E. Witten, hep-th/0106109.

\bibitem{Caldwell2003} R.R. Caldwell, M. Doran, C.M. M\"uller, G. Sch\"afer, 
C. Wetterich, astro-ph/0302505.

\bibitem{Hu1995} W. Hu, N. Sugiyama, Astrophys. J.  {\bf 444}, 489 (1995).

\bibitem{Hu1997} W. Hu, N. Sugiyama, J. Silk, Nature {\bf 386}, 37 (1997).

\bibitem{Hu2001} W. Hu, M. Fukugita, M. Zaldarriaga, M. Tegmark, Astrophys.
 J. {\bf 549}, 669 (2001).

\bibitem{Doran2} M. Doran, M. Lilley, Mon. Not. Roy. Astron. Soc. {\bf 330},
 965 (2002).

\bibitem{BOOM1} J.E. Ruhl {\it et al}, { astro-ph/0212229}.



\bibitem{Doran4} M. Doran, J. Schwindt, C. Wetterich, Phys. Rev. {\bf D64},
 123520 (2001). 




\bibitem{Bardeen} J.M. Bardeen, J.R. Bond, N. Kaiser, A.S. Szalay,
 Astrophys. J. {\bf 304}, 15  (1986).

\bibitem{Sugiyama} N. Sugiyama, Astrophys. J. Suppl. {\bf 100}, 281 (1995).

\bibitem{Bunn} E.F. Bunn, M. White, Astrophys. J. {\bf 480}, 6 (1997).

\bibitem{Louise} L.M. Griffiths, A Liddle, astro-ph/0101149.




\bibitem{Bahcall} N.A. Bahcall et al.,  Astrophys. J.{\bf 585}, 182 (2003).

\bibitem{Reiprich} T.H. Reiprich and H. B\"ohringer,  Astrophys. J.
 {\bf 567}, 716 (2002).

\bibitem{Allen} S.W. Allen, R.W. Schmidt, A.C. Fabian and H. Ebeling,
  Mon. Not. Roy. Astron. Soc.
 {\bf 334}, L11 (2002).

\bibitem{Seljak} U. Seljak, 2002, MNRAS {\bf 337}, 769 (2002).


\bibitem{Viana} P.T.P. Viana, R.C. Nichol and A.R. Liddle,  Astrophys. J.
{\bf 569}, L75 (2002).




\bibitem{Bean} R. Bean, S. H. Hansen, A. Melchiorri, Phys. Rev. {\bf D64},
 103508 (2001).

\bibitem{Hu1996} W. Hu, N. Sugiyama, Astrophys. J. {\bf 471}, 30 (1996).

\end{thebibliography}
\end{document}